\documentclass[hidelinks, fleqn,12pt]{article}
\usepackage{setspace,amsmath,amssymb,amsthm,mathrsfs,sectsty,hyperref,xcolor,graphicx,geometry,braket,epstopdf,authblk,rotating,lscape}
\usepackage[square,sort,comma,numbers]{natbib}
\hypersetup{
    colorlinks,
    linkcolor={red!50!black},
    citecolor={red!50!black},
    urlcolor={blue!50!green}
}
\allsectionsfont{\centering}
\begin{document}

\title{Statistical Analysis of Bus Networks in India}
\author{\small{Atanu Chatterjee}\thanks{atanu@wpi.edu}}
\author{Manju Manohar\thanks{manjum113@gmail.com}}
\author{Gitakrishnan Ramadurai\thanks{gitakrishnan@iitm.ac.in}}
\affil{Department of Physics\authorcr
Worcester Polytechnic Institute\authorcr
100 Institute Road, Worcester, MA 01609, USA}
\affil{Department of Civil Engineering\authorcr
Indian Institute of Technology Madras\authorcr
Chennai-600036, India}

\date{}

\maketitle

\begin{abstract}
\noindent  In this paper, we model the bus networks of six major Indian cities as graphs in \textit{L}-space, and evaluate their various statistical properties. While airline and railway networks have been extensively studied, a comprehensive study on the structure and growth of bus networks is lacking. In India, where bus transport plays an important role in day-to-day commutation, it is of significant interest to analyze its topological structure and answer basic questions on its evolution, growth, robustness and resiliency. Although the common feature of small-world property is observed, our analysis reveals a wide spectrum of network topologies arising due to significant variation in the degree-distribution patterns in the networks. We also observe that these networks although, robust and resilient to random attacks are particularly degree-sensitive. Unlike real-world networks, like Internet, WWW and airline, which are virtual, bus networks are physically constrained. The presence of various geographical and economic constraints allow these networks to evolve over time. Our findings therefore, throw light on the evolution of such geographically and socio-economically constrained networks which will help us in designing more efficient networks in the future. 
\end{abstract}

\section{Introduction}

From the neural architecture of the brain to the patterns of social interactions, many physical systems and real-world phenomena are being formulated as network models \cite{barabasi1999emergence, albert2002statistical, albert1999internet, dorogovtsev2002pseudofractal, newman2003structure, watts1998collective, ravasz2003hierarchical, niwa2003power}. These models are complex because of their size and the various emergent properties that arise due to their inter-nodal connections. Any physical, chemical, biological or social system can be visualized as a complex network; the constituting elements are known as nodes, and the interactions between them identified as links. Based on the nature of the links, these networks can be broadly classified into virtual and spatial networks. In the former category, the links are physically absent, e.g., social networks or collaboration networks, whereas, in the latter case, the links are physically present, i.e., geographically embedded road or railway networks \cite{jiang2010study, porta2006network, sen2003small, chatterjee2014scaling}. In between these two broad classes there exist networks in which the links although physically absent are, however, geographically constrained. The structure of the real-world networks such as bus or electric power grid are dependent upon the structure of the physically constrained, geographically embedded networks on which they grow and evolve. 

In the field of transportation science, the use of networks to understand the flow of entities including vehicles, cargo and pedestrians, has a long history. This traditional network flow formulation has answered many interesting engineering questions related to optimality of cost, maximality of flows and the classical, shortest path determination \cite{ahuja1988network, bertsimas2003robust}. But there exist questions that deal with the topological structure of the network, which are primarily concerned with the inter-nodal connectivity and evolution of the network, which the traditional formulation fails to address. In order to answer interesting questions, such as estimating the importance of a particular node in a network, identifying existence of hubs, analyzing the pattern of variation in shortest paths with the network size, or the robustness and resiliency of the network, we need to look at the statistical and topological properties of the network. 

Mathematically, a network is a graph, $G$, characterized by the presence of nodes, $N$, and links, $L$, connecting the nodes, such that $G=(N, L)$ where the set of nodes belong to the Euclidean space of two or three dimensions. Specific to public transit, networks are often modeled either in $L$-space or in $P$-space \cite{derrible2009network, 6950806}. In both the configurations, the nodes remain the same, for example, bus stops, metro or railway stations, whereas the pattern of the link connectivity changes. In $L$-space formulation, each pair of consecutive neighboring nodes lying along a route is considered to be connected by a link, whereas in $P$-space formulation, every possible pair of nodes belonging to a route are connected by a link. Thus, $L$-space configuration helps in understanding the relationship between the stops or nodes in general, and $P$-space helps in studying the transfers between different routes in the network. 

For each node in the set of nodes, $N=\{n_1 , n_2 , n_3 ...n_i|i\in\mathcal{I}, \forall n_i \in \mathcal{R}^n\}$, we identify the degree of a node, $k_i$ as the number of links to which that particular node is connected to. The pattern of the inter-nodal connectivity, specifically the degree-distribution, $P(k)$, of the nodes, leads to the emergence of several interesting properties of the network. Based on the degree-distribution of nodes, two prominent network models have been identified: \textbf{a}) the random network model and \textbf{b}) the scale-free network model. The random network model was first studied by Erd{\"o}s and R{\'e}nyi, and they provided two generative models where either the number of nodes and edges are fixed or each node is associated with some probability \cite{erd6s1960evolution}. Although the Erd{\"o}s-R{\'e}nyi random graph is an important model for comparison purposes, it fails to capture the essence of real-world networks, such as presence of clusters, communities, and the small-world phenomena. A more interesting model was proposed by Watts and Strogatz (WS) in order to understand real-world networks in greater depth, which is commonly known as the small-world network \cite{watts1998collective, strogatz2001exploring}. However, it has been observed that most of the real-world networks show a heavy-tailed degree distribution where the degree of few of the nodes significantly exceeds the average degree of the nodes in the network. This inhomogeneity in degree-distribution often gives rise to striking properties in the network, that has been extensively studied by Barab{\'a}si and Albert (BA) \cite{barabasi1999emergence, albert2002statistical, albert1999internet}. Both BA and WS models advocate the small-world phenomenon, which is a characteristic feature of real-world networks, such as electric power grids, WWW, Internet, social-networks, protein-yeast (metabolite) interaction networks, citation networks, and movie-actors collaboration networks \cite{barabasi1999emergence, albert2002statistical, albert1999internet, newman2003structure, watts1998collective, albert2004structural, bork2004protein, jeong2001lethality, easley2010networks, i2003least}. 

Interestingly, the above mentioned properties have been reported in various public transit networks as well \cite{derrible2009network, 6950806, sen2003small, bagler2008analysis, von2009public, woolley2011complexity, guimera2005worldwide, sienkiewicz2005statistical, angeloudis2006large}. The small-world phenomenon in transportation networks is expected since transportation facilities in a city are planned to provide maximum convenience by allowing travel between places in minimum possible time. Most transportation networks are pre-planned networks where the initial design of the network decides the presence of hubs. Transportation networks are not as large as social-networks or the Internet, and are subjected to geographical as well as socio-economical constraints. Studies on public transit networks for different cities around the world (inclusive of all modes: buses, trams, metros and monorails) have been shown to exhibit scale-free behaviour with varying values of the power-law exponent, $\gamma$ \cite{derrible2009network, guimera2005worldwide, sienkiewicz2005statistical, angeloudis2006large}. Airline and metro-networks show scale-free degree distribution patterns whereas degree-distribution in bus and rail networks tend more towards exponential patterns. The reason for this contrasting behaviour could be attributed to the two following observations: (\textbf{i}) airline-networks are not bounded by geographical constraints and (\textbf{ii}) metro-networks are \emph{local} often catering to a part of the city whereas, bus and railway-networks are \emph{global} as they are spread throughout the entire state and sometimes across the entire country. Specific to Indian scenarios exhaustive studies on public transit networks as a whole are yet to be conducted. Previous work have shown that the pattern of nodal connectivity of the Indian Railway Network (IRN) drastically differs from that of the Airport Network of India (ANI) \cite{sen2003small, bagler2008analysis}. The nature of Indian bus networks still remains understudied. 

Bus transport networks have been studied elsewhere. Analysis of the statistical properties of bus transport networks (BTNs) in China revealed their scale-free degree distribution and small-world properties. The presence of nontrivial clustering indicated a hierarchical and modular structure in the BTN. Weighted analysis of the network was done considering routes as nodes and weights as the number of common stations between the routes. The weight distribution followed a heavy tailed power law, and the strength and degree were linearly dependent \cite{xu2007scaling}. In another study, an empirical investigation was conducted on the bus transport networks (BTNs) of four major cities of China. When analyzed using $P$-space topology, the degree distribution had exponential distribution, indicating a tendency for random attachment of the nodes. The authors also evaluated two statistical properties of BTNs, viz., the distribution of number of stops in a bus route ($S$) and the number of bus routes a stop joins ($R$). While the former had an exponential functional form, the latter had asymmetric unimodal functional forms \cite{chen2007study}. The statistical analysis of the urban public bus networks of two Chinese cities, Beijing and Chengdu revealed scale free topology and small world characteristics. Presence of more hubs in the Beijing network led to a comparatively smaller exponent of degree distribution and larger clustering coefficient. Similar location of bus stops in the two cities has led to a hierarchical structure, denoted by power law behaviour (with nearly same exponents) of the weights characterizing the passenger flows \cite{ma2011power}. The rail (RTS) and bus transportation systems (BUS) in Singapore were studied with respect to their topological as well as dynamic perspectives. The stations in RTS had high average degree indicating high connectivity amongst them, while the BUS had a small average degree. Both networks had an exponential degree distribution indicative of randomly evolved connectivity. Strength of nodes defined as the sum of weight of incident edges, appeared scale free for both networks indicating the existence of high traffic hubs. The BUS network exhibited small world characteristics and had a hierarchical star like topology. RTS had slightly negative topological assortativity, while the weighted BUS displayed disassortative nature \cite{soh2010weighted}. An extended space (ES) model with information on geographical location of bus stations and routes was used to analyze the spatial characteristics of bus transport networks (BTNs) in China \cite{yang2014study}. The ES model consisted of directed weighted variations of the $L$- and $P$-space networks designated as ESL and ESP networks respectively, and the symmetry-weighted ESW network that stored information of the short-distance station pairs (SSPs). Often, two bus stations which are geographically close to each other may not have any direct bus route link between them. Such stations which are at walkable distances from each other, are defined as SSPs. The SSPs greatly influence the BTNs by reducing the transfer times as well as the number of bus routes. The average clustering coefficient of the ESW networks was considerably large, denoting a nearly circular location of the SSPs around a station. Majority of the route sections in the bus routes were short, while a few route sections connecting cities downtowns and satellite towns or special purpose BRT routes were long, leading to a power law edge length distribution of the ESL networks. 

Majority of the above studies have looked into the structural properties of the bus networks in both $L$- and $P$-spaces. The ESW network is one such network which has looked into the aspect of network redundancy due to geographical placement of the nodes. In this paper, we do a comparative study of the bus networks of some of the major Indian cities, namely Ahmedabad (ABN), Chennai (CBN), Delhi (DBN), Hyderabad (HBN), Kolkata (KBN) and Mumbai (MBN). In order to understand the structure of bus networks in India we calculate various metrics, such as clustering coefficients, characteristic path lengths, degree-distribution and assortativity. We also simulate network robustness and resiliency by first removing nodes at random, followed by targeted removal based on degree, closeness, and betweenness. This provides us with interesting results on network (nodal) redundancy, as well as structural invariance. It may seem at first that the complexity of a bus transportation network is much lesser than that of other large-scale networks, however it is the nature of the growth and the penetrative effect of these networks that makes them not only complex but interesting and worthwhile to investigate.

\section{Methodology}

We obtain the route data for all the bus networks from the respective state government websites. Every stop is considered a node, and the routes joining the stops form the set of links. We define a graph, $G = (N, L)$ where the set $N = (n_1 , n_2 , n_3 , ...)$ with each $n_i$ as a bus-stop, and the set $L = (l_1 , l_2 , l_3 , ...)$ where each $l_i$ connects the node pair $(n_i , n_j )$. The set of nodes belong to the n-dimensional Euclidean space, $\mathcal{R}^n$, and the set of links form the Cartesian product over $\mathcal{R}^n$. We define the set of routes as the set $R$ such that $\cup_i l_i\in R$ for some $i$. In order to analyze the networks, we generate the graph adjacency matrix, $A_{ij}$ such that any matrix element $a_{ij}$ of $A_{ij}$ is either equal to one or zero depending upon the existence of a connecting link between node-pair $(i, j)$. The degree of any node is given as $k_i = \Sigma_i a_{ij}$. The above formulation generates a $L$-space network without weights. In order to assign weights, we calculate the route overlaps between a pair of nodes which we call edge-weights, $w_{ij}$. The degree strength matrix is given by $s_{ij} = a_{ij} \times w_{ij}$ and the weighted degree or node-strength as $s_i = \Sigma_i s_{ij}$. Since the flow of transport is along both the directions, we consider the network links to be undirected. The local clustering coefficient is given by $C (i) =\frac{2|a_{ij}: (n_i , n_j)\in N, a_{ij}\in A_{ij}|}{k_i (k_i - 1)}$ where $a_{ij}$ is the link connecting node pair $(i, j)$, and $k_i$ are the neighbours of the node $n_i$. The neighbourhood, $n_i$, for a node, $i$ is defined as the set of its immediately connected neighbours, as $n_i = \{n_j : l_i \in L \wedge l_j \in L\}$. For the complete network, Watts and Strogatz defined a global clustering coefficient \cite{newman2003structure, watts1998collective}, $C =\Sigma_i C_i /n$. The weighted clustering coefficient is given as \cite{barrat2004architecture} $C^w (i) = \frac{1}{s_i (k_i - 1)}\Sigma_{j, h}\frac{w_{ij} + w_{ih}}{2}a_{ij}a_{ih}a_{jh}$. 

Another important measure is the characteristic path length, $l_{ij}$ which is defined as the average number of nodes crossed along the shortest paths for all possible pairs of network nodes. The average distance from a certain vertex to every other vertex is given by $d_i = \Sigma_{i\neq j}\frac{d_{ij}}{|N(G)|-1}$. Then, $l_{ij}$ is calculated by taking the median of all the calculated $d_i$ $\forall i \in \mathcal{R}^n$. In order to check the small-world property, we generate random graphs of same size, i.e., keeping network size $N$ constant. However, the network topology of a random graph is governed by a wiring probability, $p_w$ which determines the connectedness of the network (or the number of edges of the network). In order to generate random networks of comparable sizes (similar number of nodes and edges), we calculate the wiring probability as $\frac{p_w N^2}{2}\sim N$. The centralities, betweenness and closeness tell us the relative importance of nodes in the network. Betweenness centrality of any node is calculated as, $C_B (i) = \Sigma_{s\neq i \neq t}\frac{\sigma_{s, t}(i)}{\sigma_{s, t}}$, where $\sigma_{s, t}$ is the number of shortest paths connecting $s$ to $t$ and $\sigma_{s ,t}(i)$ number of shortest paths connecting $s$ to $t$ but passing through $i$. Likewise, closeness centrality for any node is calculated by $C_C (i) = \Sigma_i \frac{1}{a_{ij}}$. The average closeness is the harmonic mean of the shortest paths from any node to every other node. In weighted networks, usually the edge weights are considered as cost functions; therefore, larger the edge weight, lesser is the node's closeness, as the cost of travel would be large. However, in our case the edge weights play an altogether different role signifying the `ease' of travel. Hence, we take the inverse of edge weights during the calculation of weighted $C_C$ as in collaboration networks given by $C^{w}_C (i) = \min\Sigma_i (\frac{1}{w_{ij}})$. 

The degree-assortativity or the Pearson correlation coefficient of degree between pairs of linked nodes is given by $\Sigma_{jk}\frac{jk(e_{jk}-q_j q_k)}{\sigma^{2}_q}$, where $e_{jk}$ is the joint probability distribution of the remaining degrees of the two vertices at either end of a randomly chosen edge with $\Sigma_{j, k} e_{jk} =1$ and $\Sigma_j e_{jk} = q_k$. Here, $q_k$ is the normalized degree-distribution of the remaining degrees, and $\sigma^{2}_q$ is the variance of the distribution $q_k$ given by\cite{newman2002assortative} $\sigma^{2}_q = \Sigma_k k^2 q_k - [\Sigma_k kq_k]^2$. The degree-distribution $P(k)$ gives the probability of finding a node with a degree $k$ in the network, which basically represents the ratio of all the nodes with degree equal to $k$ to the size of the network, $N$. The degree-distribution is observed to follow a heavy-tailed function. The equation for the power-law or exponential fits (in Table 1 and Figure 2) are calculated using Maximum Likelihood Estimation (MLE) and the Kolmogorov-Smirnov test is employed to check for goodness of fit \cite{clauset2009power}. The degree-strength correlation is evaluated using linear-regression model, and the least-square error is calculated. 

\section{Results}

The datasets were obtained from the government websites of Ahmedabad BRTS (ABN), MTC (CBN), DTC (DBN), APSRTC (HBN), CSTC (KBN) and BEST (MBN). In Figure 1, we plot the network structure using force directed algorithms. The figure compares the structural construct of the networks. We can clearly observe the nature of connectivity between the nodes in the different networks. While DBN is densely packed, CBN, HBN and KBN are sparse. The network structure of MBN is particularly striking. The long branches with multiple intermediate nodes as seen from the figure cause the characteristic path-length, $l_{ij}$ of MBN to increase abnormally (see Table 1). We also calculate the modularity of the networks to identify community structure. Networks with high modularity have dense connections between the nodes within the same modularity class but weak connections between nodes in different modularity class. In order to identify communities we colour-code the nodes based upon the modularity classes. Community detection in bus networks help us in identifying the different zones of operation. As large as six communities were identified for CBN and MBN whereas fewer (four or less) communities were identified for ABN, DBN, HBN and KBN. 

In Table 1, we present the statistical analysis for the various networks in a tabular form. It can be seen from the table that the network sizes of all the cities are comparable to each other, except that of KBN because CSTC is localized and operates as a subdivision of West Bengal Surface Transport Corporation (WBSTC) that operates buses in the entire state. The network density, $\rho$, which is the ratio of the number of edges in a given network to the corresponding completely connected graph varies from $0.001$ to $0.006$. An interesting feature is the variation of the characteristic path length $l_{ij}$ from as low as $3.87$ to as high as $10.02$. In order to get a deeper insight into the structure of these networks, we carried out a weighted analysis by assigning a weight corresponding to the overlap of routes connecting a particular pair of nodes that helps us understand the potential flow of traffic between that nodal pair. The weighted degree of a node or its strength is observed to follow a heavy-tailed distribution on a double logarithmic scale, and the node strength and node degree are found to be related non-linearly. This implies that the potential traffic at a node due to route overlaps increases exponentially as compared to the actual number of routes it is connected to \cite{barrat2004architecture}. 

We observe that the average clustering coefficient, $C_{av}$ also shows a remarkable variation from $0.07$ to as high as $0.26$. We check the presence of small-world phenomenon in the above networks by generating random graphs with the same number of nodes and comparable number of edges, and calculate the characteristic path length, $l_{ij}^{rand}$ and average clustering coefficient, $C_{av}^{rand}$ in each case. Upon comparing with the data in Table 1, we find that $C_{av} >> C_{av}^{rand}$ each time, whereas $l_{ij}$ is either comparable to $l_{ij}^{rand}$ or $l_{ij} < l_{ij}^{rand}$. Based upon the above comparisons, we can state that the bus networks show small-world phenomenon. As we discussed earlier, $L$-space formulation merely gives the relationship between bus stops and bus routes, whereas it is the $P$-space formulation which helps in determining the number of transfers, or in this case, number of bus changes. We can estimate the number of bus changes required by looking at the average number of bus stops present in each of the routes. CBN and MBN typically show the largest magnitudes of characteristic path-lengths in $L$-space. A $P$-space analysis for both CBN and MBN, reveals the number of transfers as low as $2-3$. Thus, all the networks studied in this paper show small-world behaviour in $P$-space topology~\cite{chatterjee2015studies}. 

\begin{table}[t]
\begin{center}
\begin{tabular}{|cccccccc|}
\hline
{\bf Bus routes} & {\bf Nodes} & {\bf Edges} & {\bf $l_{ij}$} & {\bf $C_{av}$} & {\bf $\gamma$} & {\bf Assortativity} & {\bf $<k>$} \\ \hline
{\bf ABN}        & 1103        & 2582        & 5.59           & 0.19                         & 2.47                      & 0.07              & 3.67           \\
{\bf CBN}        & 1644        & 2732        & 9.02           & 0.142                         & 3.05                       & 0.09              & 3.31          \\
{\bf DBN}        & 1557        & 4287        & 5.51           & 0.18                         & 3.13                 & 0.07              & 9.88           \\
{\bf HBN}        & 1088        & 2954        & 3.87           & 0.26                         & 3.52                      & -0.03             & 23.88          \\
{\bf KBN}        & 518         & 884         & 5.72           & 0.08                         & 4.96                     & -0.01             & 6.72           \\
{\bf MBN}        & 3131        & 6443        & 10.02          & 0.18                         & 3.25                 & 0.45              & 33.38          \\ \hline
\end{tabular}
\caption{Tabular representation of the statistical data for the bus routes of six major Indian cities ($l_{ij}$ = characteristic path length, $C_{av}$ = average weighted clustering coefficient, $\gamma$ = power-law exponent, and $<k>$ = average node degree).}
\end{center}
\end{table}

\begin{sidewaysfigure}[hb]
\begin{center}
\includegraphics[width=1\linewidth]{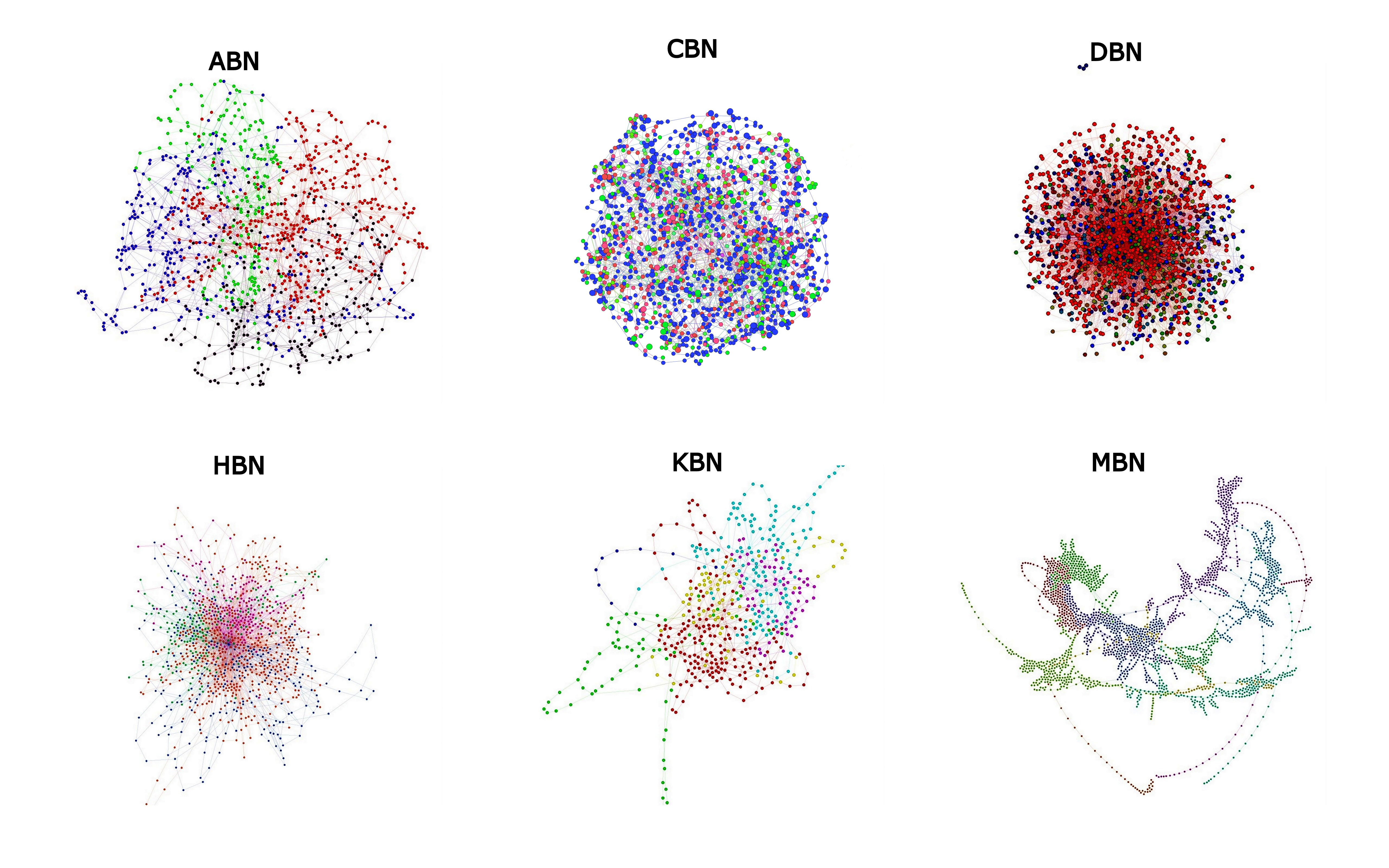}
\caption{Figure shows the network structure of the different bus routes where each node represents a bus stop. The plots are generated using force directed algorithms and the colour of nodes partition the networks into different communities. Where ABN, DBN and HBN show typical scale-free structure observe the long routes present in CBN and MBN.}
\end{center}
\end{sidewaysfigure}

As discussed earlier, node-degree distribution plays an important role in understanding the structure and evolution of complex networks. In Figure 2 (a), we plot the degree distribution for all the networks on a double logarithmic scale. The degree-distribution patterns show mostly heavy-tailed characteristics, with MBN showing a slight deviation from the power-law behaviour. In Figure 2 (b), we plot the centrality distributions (closeness and betweenness), $P(C_C)$ and $P(C_B)$ in the first two rows for ABN, HBN (scale-free) and MBN (non scale-free) on a double logarithmic scale to contrast the differences between scale-free networks and non scale-free ones. We find that the distribution function follows an exponential decay given by $P(C_C) \sim \exp(-\lambda C_C)$ (similarly for $C_B$) where the value of the exponent $\lambda$ is shown in each of the plots. In the last row, we plot the variation of betweenness centrality with the degree of a node which follows a power-law relationship, given as $C_B\sim k^\alpha$ with the magnitude of the exponent $\alpha$ also shown in the plots. 

\begin{figure}[hb]
\begin{center}
\includegraphics[width=1\linewidth]{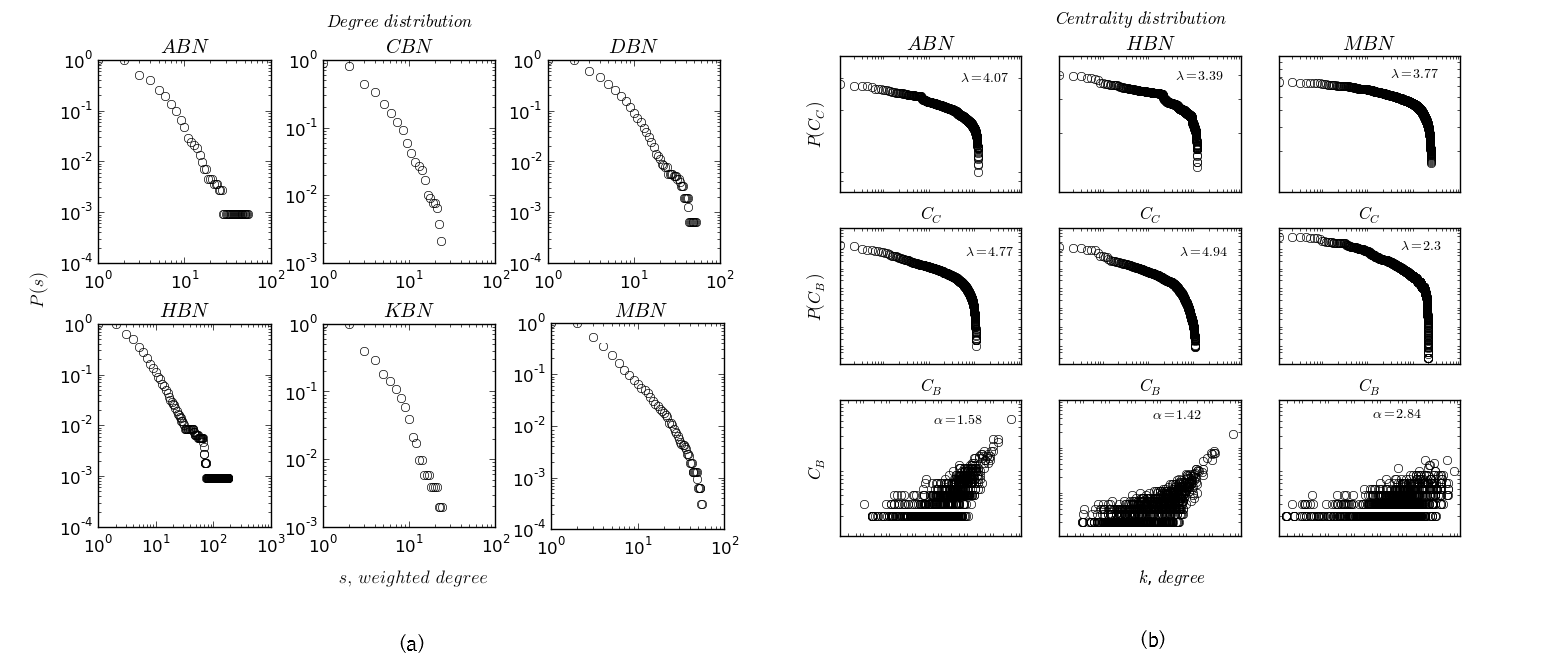}
\caption{(a) Figure shows the degree-distribution, $P(k)$ on a double logarithmic scale; (b) Figure shows centrality distribution for betweenness ($C_B$) and closeness centralities ($C_C$) with the decay exponent $\lambda$ (inset). The plots in the last row show degree-betweenness dependency with exponent $\alpha$ (inset).}
\end{center}
\end{figure}

In Figure 3, we plot the response of the network\rq{}s characteristic path length, $l_{ij}$ to random and systematic perturbation. We simulate the robustness and resiliency of the networks by modeling perturbations as node removals. Due to their strong assortative nature, MBN and CBN disintegrate into separate entities very quickly, whereas the other networks remain connected upto atleast $4\%$ of node removals. It is observed that in all the cases the targeted node removals are crucial for the network to remain connected. In the regime of $p_i \leq 4\%$, a closer look reveals that the magnitude of $\textbf{l}_{ij}$ does not change much (at most it increases by one `hop'). Finally, in Figure 4, we plot the degree-distribution for ABN and MBN after removing $20\%, 40\%$, and $60\%$ of the nodes in order to check the invariance in the topological structure of these networks. We choose ABN and MBN as these two networks are topologically different. 

\begin{sidewaysfigure}
\begin{center}
\includegraphics[width=0.8\linewidth]{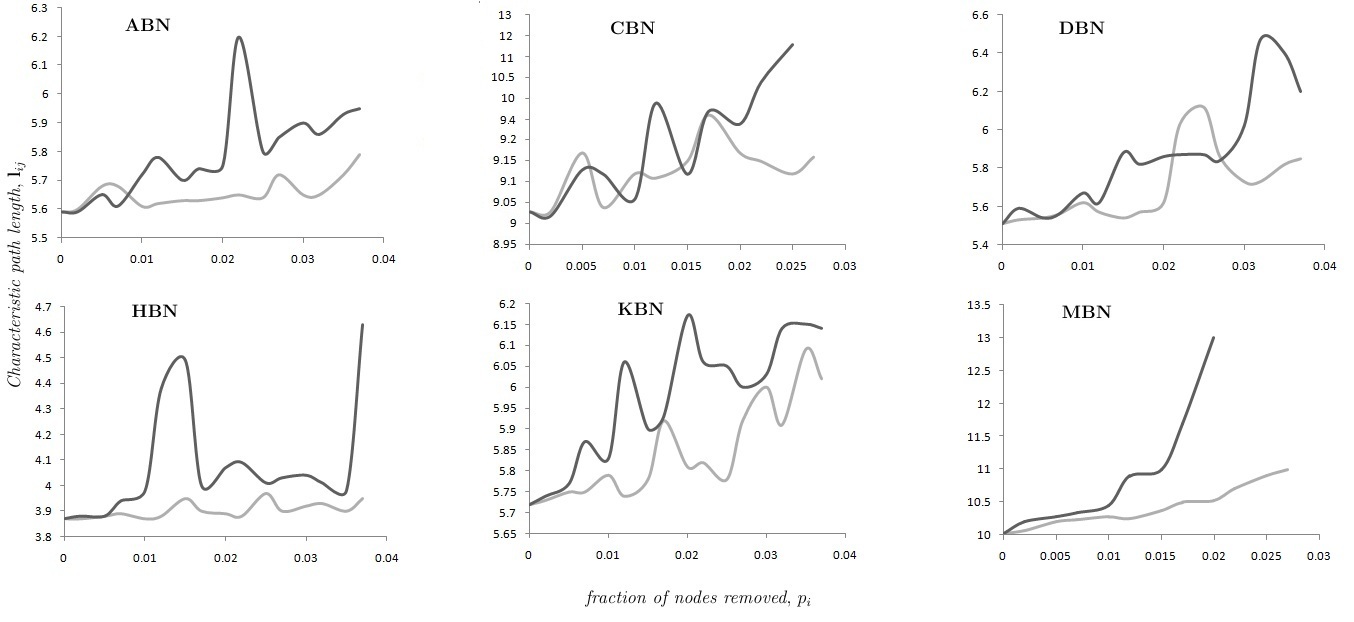}
\caption{Figure shows the variation in $l_{ij}$ with network size upon random and targeted node removals. The dark line represents degree-based node removals, and the light line represents random node removals. The X-axis represents the fraction of nodes removed.}
\end{center}
\end{sidewaysfigure}

\begin{sidewaysfigure}
\begin{center}
\includegraphics[width=1\linewidth]{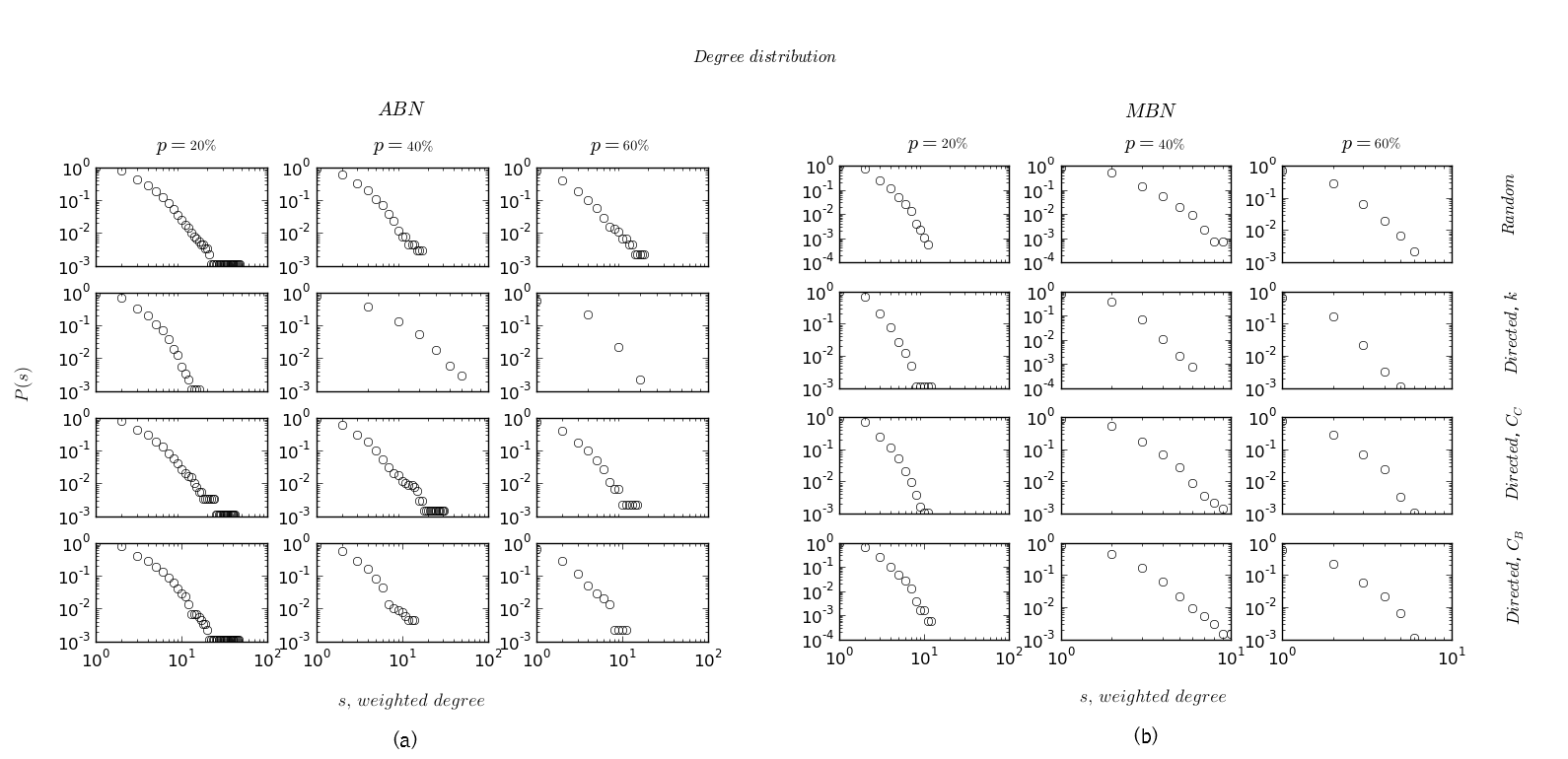}
\caption{Figure shows degree-distribution plots for (a) ABN and (b) MBN subjected to random and directed attacks for percentage of node removals, $p = 20\%, 40\%$ and $60\%$.}
\end{center}
\end{sidewaysfigure}

\section{Discussion}

In this paper, we analyzed the statistical properties of the bus routes of the six Indian cities, namely Ahmedabad, Chennai, Delhi, Hyderabad, Kolkata, and Mumbai. Our analysis suggests that the bus networks show a wide spectrum of topological structure from power-law to exponential with varying magnitude of the power-law exponent $\gamma$. Ahmedabad (ABN) is particularly interesting in this regard because it has a BRTS (Bus Rapid Transit System) with dedicated lanes - a type of public transit system that is yet to be introduced at a large scale in India. ABN's BRTS, thus, holds a structural advantage by the presence of many hubs to which extreme routes are connected, a structure similar to WWW or the airline networks (WAN and ANI) \cite{barrat2004architecture, bagler2008analysis}. As we saw in the earlier sections, CBN and MBN do not show the small-world property in $L$-space. They, however, do show the small-world property in terms of transfers ($P$-space topology), as majority of the places can be visited by making as little as 2 to 3 bus changes~\cite{chatterjee2015studies}. The structural relationship between bus stops as observed from the degree-distribution plots in Figure 2 is of particular interest. In Figure 2, we plot the weighted degree-distribution of the networks which capture the strength of the nodes with respect to the traffic handled in terms of the number of routes. In order to check for correlations between node degree, $k$ and node weighted-degree, $s$ we plot them on a double-logarithmic scale. Interestingly, ABN shows a strong correlation as, $s\sim  k^{\beta}$ with $\beta = 1.27$ and $R^2 = 0.91$, whereas the other networks fail to show such strong relationships (CBN, KBN and HBN show similar relationships with $\beta \sim 1.44 - 2.08$, however, with lower correlation coefficients, $R^2 \sim 0.60 - 0.74$). The degree-distribution in case of ABN has the power-law exponent, $\gamma$ as $2.47$, whereas the degree-strength exponent, $\beta$ is found to be $1.27$. This implies that the strength of a node increases faster as compared to its degree indicating a sense of order in ABN where higher degree nodes, for example, large or important bus stops, handle heavy traffic as majority of the routes pass through them. This is definitely missing in the other networks where the edge weights or routes seem to be more randomly distributed. Also the topological structure of the road networks in the city of Ahmedabad show a scale-free degree distribution with $\gamma = 2.5$ and $l_{ij} = 5.20$, which is very similar to ABN \cite{porta2006network} (see Table 1). 

In Figure 2 (b), we plot the centrality distribution for betweenness ($C_B$) and closeness ($C_C$). We consider betweenness and closeness because they play a crucial role from a transportation perspective. $C_C$ is a measure of a node's relative importance in the network due to the existence of shortest paths from that particular node to every other node in the entire network. $C_B$ on the other hand acts as a bridging node connecting different parts of the network together. When traveling from one node to the other, it is often beneficial to get to the node with the highest value of $C_C$ first if a direct path does not exist between the origin-destination pair. Often transportation network of a city is planned in a way such that the hubs allow maximum number of routes to pass through them, and all other nodes in the network to be easily reachable from them. Since, centrality is positively correlated to node degree, the hubs in a network also tend to have the largest degrees. We found this pattern in all the networks, ($C_B \sim k^\alpha$); however, in DBN and MBN the relationship between degree and centrality is not that strong perhaps due to the presence of \emph{noise} in the network due to random attachment of nodes (see Figure 2 (b) last row). The noise or the presence of redundant nodes (links) due to random attachment of the nodes in the network causes the degree-distribution patterns to shift from a purely power-law decay to truncated power-law and exponential decays. The presence of these redundant nodes increase the degree of non-central nodes which is observed in the degree-centrality plots (see Figure 2). These nodes due to their random placement tend to appear at random places in the network causing hindrance in the direct connectivity of the hubs. The networks (except CBN and MBN) therefore show disassortative or weakly assortative behaviours. We also observe that the centrality-distribution functions follow exponential decay, as $P(C_C)\sim \exp(-\lambda C_C)$ (similarly, for $C_B$) which shows that nodes in a network are different, i.e., some nodes are more `central' as compared to other nodes. An interesting observation is that nodes in the networks tend to connect to existing high degree nodes preferentially whereas such a preferential attachment rule is missing when, for example, node-betweenness is considered as the metric. A close observation in Figure 2 (b) reveals that nodes with high betweenness certainly have high degrees however, the reverse is not true. 

Some nodes do not play any significant role in the network's overall functionality, i.e., they are redundant. In Figure 3, we evaluate the network's response to external perturbations by random and directed removal of nodes. We fix an important measure $l_{ij}$ and check its variation upon percentage removal of nodes (bus stops). As we saw earlier, CBN and MBN due to their strong assortative behaviour, seem to be very sensitive to node removals as they quickly disintegrate, whereas ABN, DBN, HBN and KBN do not show any significant change in $l_{ij}$ upto $4\%$ of node removal. This basically amounts to roughly $40 - 70$ nodal redundancy (in numbers), that if removed can reduce cost of construction, operation, and maintenance significantly in the network. However accessibility for all users has to be carefully studied before removing any node. We also observe that the clustering coefficient $C$ varies inversely with the node degree which implies that the nodes with low clustering coefficients tend to have higher degrees and vice-versa. This is because nodes (bus stops) having higher degree will be a part of multiple bus routes whereas, those bus stops through which fewer bus routes pass will have lower degree. Thus, it is more likely for the nodes in the later case to form clusters as compared to the ones which are connected to multiple bus routes.

Finally in Figure 4, we observe that the topological structure of the networks are preserved (Figure 4 (a) and 4 (b)) when the networks are subjected to large number of node removals. It can also be clearly seen that the networks are degree-sensitive. Degree-biased node removal causes the heavy tails in the degree-topology to disappear thus signifying gradual decrease in the number of hubs. Interestingly, a similar effect is also observed when the nodes are removed based upon their betweenness centralities. Although, the effect is relatively less significant, it is more when compared to closeness biased node removals and random node removals. In Figure 4 (b), we plot the degree-distributions for MBN with respect to percentage node removal. In case of MBN it is particularly interesting to note that the degree-distribution plots, which originally showed a better fit for exponential distribution (Figure 2 (a)), evolves into a scale-free topology (as can be observed from straight line slope in the double-logarithmic scale) with varying power-law exponent, $\gamma$, when nodes are removed. At $20\%$ node-removal, MBN starts showing heavy-tailed degree topology. The above phenomenon could be attributed to the reduction of noise (randomness of connectivity and nodal redundancy) due to removal of nodes. 

Also, from Table 1, we observe that the bus networks, like all other surface transport networks are assortative in nature with HBN and KBN showing weak disassortative behaviour. The strong assortativity observed in these networks result in increased characteristic path-lengths. Since, the nodes (bus stops) are spatially distributed throughout the city, the tendency of similar nodes to attach to nodes with similar statistical properties causes the characteristic path-lengths to increase significantly. From a transportation perspective, assortative mixing is beneficial as this will allow direct connectivity between hubs. However, it will also increase the number of hops in traversing from any given source to a destination within the network. In terms of transfers, the small-world property is retained, yet the traveling time between any random origin-destination pair will increase, due to delays associated with numerous intermediate stops. 

As noted earlier, bus networks form a specific class of complex networks that grow and evolve over physically constrained spatial networks. Road intersections are usually separated by a distance which is geographically much smaller as compared to the distance between bus stops; therefore, our results emphasize that transportation undoubtedly brings the world closer. What we observed from our paper is that bus networks show scale-free topology and small-world property in the number of transfers.  Also, from the above analysis we observe that the bus networks although structurally different, show similar as well as self-similar topological structures. With the exception of MBN, all the networks show scale-free topology with MBN showing slight deviation towards an exponential distribution. The presence of heavy-tails in the degree-distribution plots imply a preferential attachment rule, the tendency of high degree nodes to cluster with low degree nodes reveal a hierarchical organization, and the stability of characteristic path length with gradual removal of nodes reveal the presence of nodal redundancy in the network. 

The present study opens before us new horizons for efficient transportation network designing and planning. Questions such as: what are the statistical properties of the network that will ensure efficiency or how network topology is related to the statistical properties and vice-versa would be both challenging and worthwhile to answer. It would be exciting to come up with innovative models to capture the growth and evolution of real-world large scale public transit networks. Developing generative methods to reduce noise (in the network) due to random node attachment by including geographic and socio-economic constraints such as demand, flow, and cost, to maximize certain network parameter(s) or node-utility function(s) based on the above constraints is another promising area of future work. 

\bibliographystyle{plos2015}

\section{Acknowledgments}

The authors acknowledge the support from Center of Excellence in Urban Transport at the Indian Institute of Technology, Madras, sponsored by the Ministry of Urban Development, Government of India and the Information Technology Research Academy, a Division of Media Labs Asia, a non-profit organization of the Department of Electronics and Information Technology, funded by the Ministry of Communications and Information Technology, Government of India.

\section{Author contributions statement}

A.C. and G.R. conceived the idea,  A.C. and M.M. collected the data and wrote the manuscript, A.C. wrote the codes and ran the simulations, A.C. and G.R. analyzed the results. All authors reviewed the manuscript. 

\section{Additional information}

\textbf{Competing financial interests} The authors declare no competing financial interests.

\end{document}